\documentclass[conference, 9pt]{IEEEtran}
\IEEEoverridecommandlockouts
\usepackage{cite}

\usepackage{amsmath,amssymb,amsfonts}
\usepackage{graphicx}
\DeclareGraphicsExtensions{.pdf,.png,.jpg}
\usepackage{textcomp}
\usepackage{xcolor}
\usepackage{booktabs}
\usepackage{array}
\usepackage{caption}
\usepackage{subcaption}
\usepackage{setspace}
\usepackage{algorithm,algpseudocode}
\usepackage{comment}
\def\BibTeX{{\rm B\kern-.05em{\sc i\kern-.025em b}\kern-.08em
    T\kern-.1667em\lower.7ex\hbox{E}\kern-.125emX}}
\begin{document}

\title{\huge \bf
FIXAR: A Fixed-Point Deep Reinforcement Learning Platform \\ with Quantization-Aware Training and Adaptive Parallelism\\
}
\author{\IEEEauthorblockN {Je Yang \hspace{1cm} Seongmin Hong \hspace{1cm} Joo-Young Kim}
\IEEEauthorblockA{{School of Electrical Engineering} \\
{KAIST}\\
\texttt{\{yangje, seongminhong, jooyoung1203\}@kaist.ac.kr}}
}

\maketitle

\begin{abstract}
\small	
Deep reinforcement learning (DRL) is a powerful technology to deal with decision-making problem in various application domains such as robotics and gaming, by allowing an agent to learn its action policy in an environment to maximize a cumulative reward. Unlike supervised models which actively use data quantization, DRL still uses the single-precision floating-point for training accuracy while it suffers from computationally intensive deep neural network (DNN) computations.
\par
In this paper, we present a deep reinforcement learning acceleration platform named FIXAR, which employs fixed-point data types and arithmetic units for the first time using a SW/HW co-design approach. We propose a quantization-aware training algorithm in fixed-point, which enables to reduce the data precision by half after a certain amount of training time without losing accuracy. We also design a FPGA accelerator that employs adaptive dataflow and parallelism to handle both inference and training operations. Its processing element has configurable datapath to efficiently support the proposed quantized-aware training. We validate our FIXAR platform, where the host CPU emulates the DRL environment and the FPGA accelerates the agent's DNN operations, by running multiple benchmarks in continuous action spaces based on a latest DRL algorithm called DDPG. Finally, the FIXAR platform achieves 25293.3 inferences per second (IPS) training throughput, which is 2.7 times higher than the CPU-GPU platform. In addition, its FPGA accelerator shows 53826.8 IPS and 2638.0 IPS/W energy efficiency, which are 5.5 times higher and 15.4 times more energy efficient than those of GPU, respectively. FIXAR also shows the best IPS throughput and energy efficiency among other state-of-the-art acceleration platforms using FPGA, even it targets one of the most complex DNN models.
\end{abstract}
\vspace{0.05in}

\vspace{-0.08in}
\section{\bf \textsc{Introduction}}
\vspace{-0.03in}
Reinforcement learning (RL) is a promising area of machine learning that studies how an agent should take actions in an environment in order to maximize a long-term cumulative reward. It aims to solve a complex decision-making problem in a setting where the agent constantly updates its action policy based on the reward feedback from the environment. Recently, deep reinforcement learning (DRL) that utilizes a deep neural network (DNN) for the action policy to train has been proposed \cite{mnih2013playing, mnih2016asynchronous, lillicrap2015continuous, DBLP:journals/corr/SchulmanWDRK17}. This deep learning approach becomes very popular like in other machine learning disciplines, as it observes widespread success in various applications such as robotics, industrial control, and game playing \cite{kempka2016vizdoom, 6787321,  zhang2015visionbased}. Unlike supervised learning requires a number of labeled input/output pairs for training of the model, DRL uses its own inference samples to train an agent to take desirable actions. However, DRL's training process is computationally expensive as it requires repeated computations of the forward and backward propagation.
\par
Data quantization is an effective technique to reduce the size of DNN models by replacing 32-bit floating-point weights and activations to less complicated representations such as lower-bit fixed-point numbers with negligible or even no accuracy loss through re-training \cite{han2015deep, zhou2016dorefa}. It is beneficial for hardware because the reduced model requires smaller memory storage as well as smaller memory bandwidth. It also enables efficient computations by employing simpler arithmetic units with low bit-precision. Due to these strong benefits, many state-of-the-art DNN platforms try to use low-bit fixed-point format over expensive floating-point format if possible\cite{yang2019quantization,jacob2018quantization}. Unfortunately, most of these quantization researches are done with supervised models, especially focused on inference. It is questionable if we can apply the same quantization techniques to deep reinforcement learning. As the agent's current decision heavily influences the future state and action, it is hard to predict that how quantization will affect the policy's long-term decision in complex environment \cite{krishnan2019quantized}.
\par
In this paper, we propose a fixed-point deep reinforcement learning acceleration platform named FIXAR. FIXAR successfully employs dynamically changing dual fixed-point data types for the first time in the training process of deep reinforcement learning. Using a SW/HW co-design approach, its FPGA based accelerator achieves the highest energy efficiency among existing state-of-the-art compute platforms.
\begin{itemize}
\item {\bf Algorithm Design}
We propose a quantization-aware training algorithm which enables to reduce the full data precision in fixed-point by half after a certain number of training steps to keep its training accuracy in DRL. Based on this algorithm, the FIXAR's hardware accelerator employs fast and energy-efficient fixed-point arithmetic units instead of expensive floating-point arithmetic units.
\item {\bf Accelerator Design} We design a FPGA based accelerator responsible for running DNN inference and training operations with supporting the proposed quantization-aware training. It is the first hardware accelerator that supports both inference and training with dual bit-precision in fixed-point for DRL. To this end, we propose the adaptive array processing core that exploits different dataflows and parallelisms to handle both forward and backward propagation. Its processing element has configurable datapath to run a quantized model seamlessly with doubling its throughput. With eliminating external memory access using an efficient on-chip memory structure, FIXAR's accelerator achieves a very high training throughput.
\par
\item {\bf Platform Design} By integrating the proposed training algorithm and accelerator design together, we implement our FIXAR platform on a CPU-FPGA system, where the host CPU emulates the environment and the FPGA accelerator runs the agent's DNN operations. We successfully validate the FIXAR platform by running multiple benchmarks in continuous action spaces from MuJoCo environment, based on a widely used DRL algorithm called DDPG.
\end{itemize}
\par
Finally, the FIXAR platform achieves 25293.3 inferences per second (IPS) throughput and 2638.0 IPS/W accelerator efficiency from system-level benchmarking, which is 2.7 times faster and 15.4 times more energy efficient than those of the CPU-GPU platform without any accuracy degradation. Among other state-of-the-art acceleration platforms using FPGA, the FIXAR shows the best IPS performance and energy efficiency even it targets one of the most complex DNN models.

\vspace{-0.05in}
\section{\bf \textsc{Background}}
\vspace{-0.05in}
\subsection{Deep Reinforcement Learning}
\begin{figure}[t!]
\centering
   \includegraphics[width=5cm]{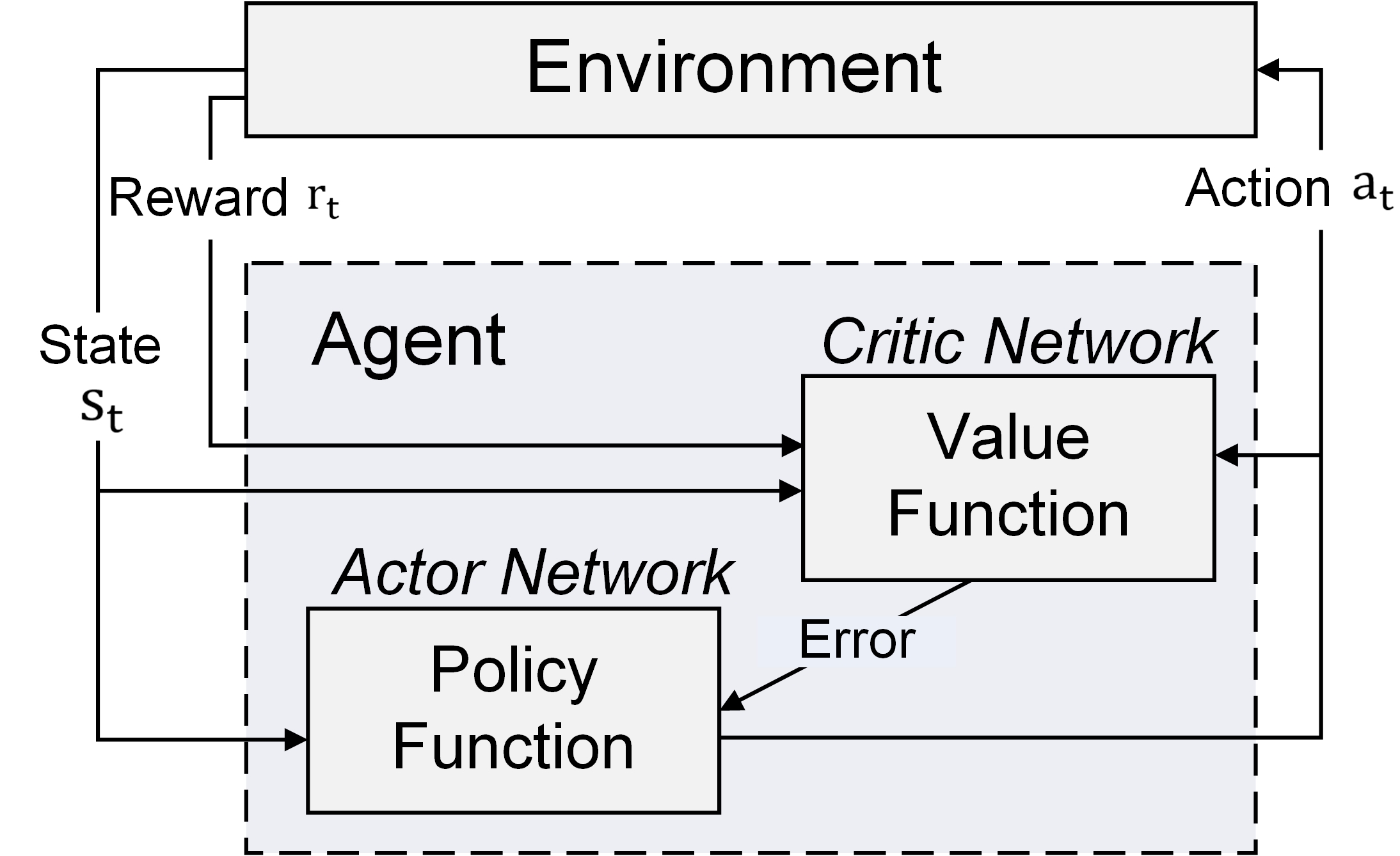}
   \hfil
\caption{Actor-critic reinforcement learning algorithm}
\label{RL_env}
\vspace{-0.17in}
\end{figure}

\vspace{-0.03in}
In reinforcement learning, an agent interacts with its environment for the aim of learning reward-maximizing behavior (Figure \ref{RL_env}). At each timestep $t$, the agent receives an state $s_t$ and selects an action $a_t$ based on its policy $\pi$. After sending the selected action $a_t$ to the environment, the agent receives a reward value $r_t$ and the next state $s_{t+1}$ from the environment. The agent continuously interacts with the environment for the goal of maximizing the cumulative reward $R_t = \sum_{i=t}^{T}\gamma^{i-t}r_i$ , where $T$ is the total timesteps for an episode and $\gamma \in (0,1]$ is a discount factor that reflects the importance of future rewards. The Q-value $Q_\pi = E_\pi[R_{t+1} + \gamma R_{t+2} + ... | S_t = s_t, A_t=a_t]$ represents how effective the action $a_t$ is in state $s_t$ at timestep $t$. In conventional reinforcement learning, Q-learning algorithm which stores the Q-values for all state-action pairs in the Q-table and recursively finds the optimal actions that maximizes the Q-value is commonly used. However, this table based training of Q-learning becomes easily unstable when the number of states/actions increases in a complex continuous environment because the chance of the agent visiting a particular state-action pair gets increasingly small. To solve this problem, actor-critic algorithm is suggested.
\par
Actor-critic algorithm uses a deep neural network approach in learning the action policy. The actor calculates the action based on the policy network $\pi(a_t|s_t;\theta)$ where $\theta$ is the network's model. On the other hand, the critic evaluates how good the selected action is based on value network $V(s_t;\theta)$. The DNN model parameters are updated based on the calculated values from the value network, in the direction where more rewards can be obtained through optimal actions. Among actor-critic algorithm, Deep Deterministic Policy Gradients (DDPG) \cite{lillicrap2015continuous} (and its variants \cite{barthmaron2018distributed, fujimoto2018addressing}) is known to be one of the best and is successfully demonstrated in continuous control domain.

\vspace{-0.09in}
\subsection{Quantization in Deep Reinforcement Learning}
\vspace{-0.03in}
DNN model quantization is an active research area in machine learning, which aims to use as smallest bit-precision as possible in order to minimize both memory storage and bandwidth requirements. Since the accuracy is the most important goal to achieve in model training, many training methods still use the standard 32-bit single-precision floating-point format to cover a wide dynamic range of calculated gradients. One of the biggest problems of a quantized format in training process is the data truncation issue caused by a limited dynamic range. For fixed-point quantization, this truncation issue gets even worse. In addition, inherent error generated by quantizing a higher-bit floating-point data to a lower-bit fixed-point representation should be handled as well. To overcome these challenges, Jain et al.\cite{jain2018compensated} introduced a new fixed-point representation that contains compensation bits to dynamically adjust the error introduced during quantization. Zhang et al. \cite{zhang2020fixed} proposed the quantization scheme that changes the bit-width automatically by layers in the fixed-point back-propagation.
\par
As the deep reinforcement learning tries to address a continuous decision-making problem, which is quite different from recognition and classification tasks normally handled by conventional supervised models, it is questionable if we can apply the same quantization methodologies. Krishnan et al. \cite{krishnan2019quantized} showed that certain reinforcement learning algorithms and environment tasks are more problematic to quantize because of their weight distributions. They also demonstrated that quantization may have a positive effect on the training as the induced noise diversifies the action exploration.

\vspace{-0.01in}
\section{\bf \textsc{FIXAR Platform}}
\vspace{-0.03in}
\begin{figure}[t]
\centering
   \includegraphics[width=7.3cm, height=8cm]{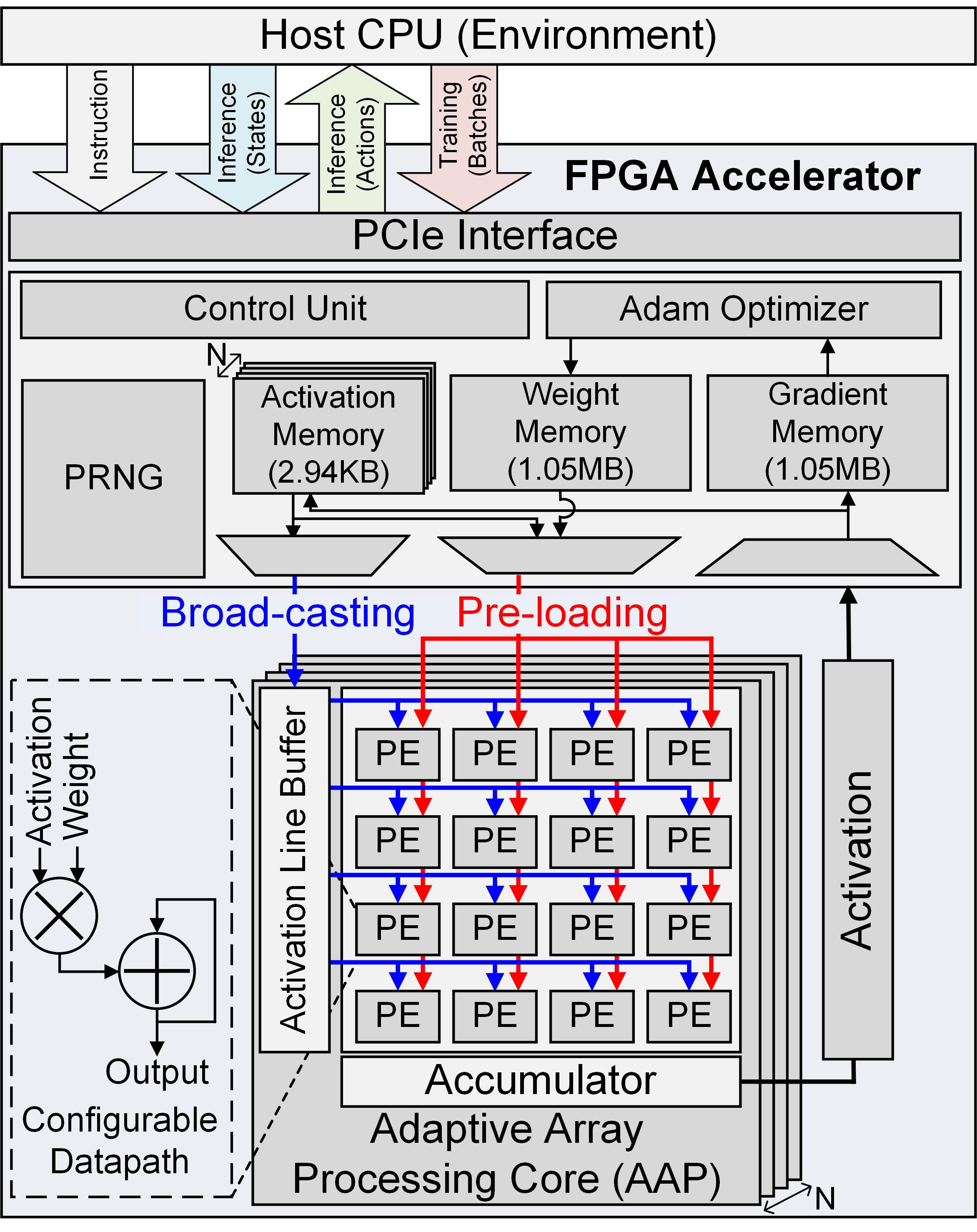}
   \hfil
\caption{Overall FIXAR platform}
\vspace{-0.2in}
\label{FIXAR}
\end{figure}
\begin{figure}[t]
\centering
   \includegraphics[width=9cm]{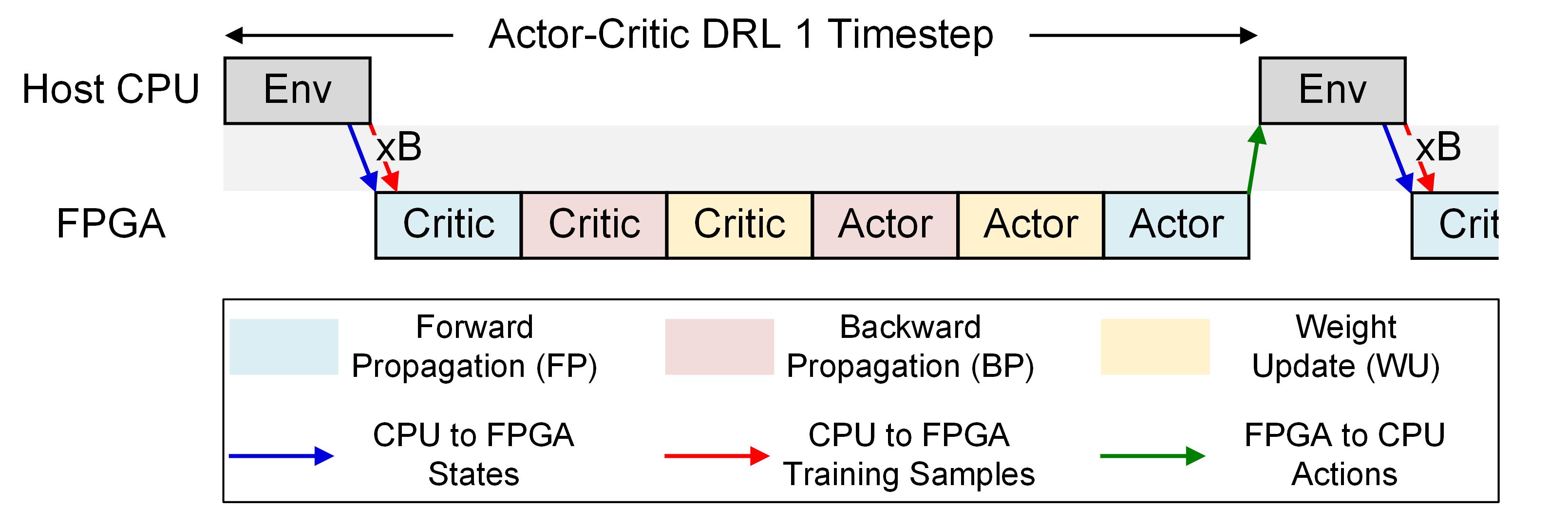}
   \hfil
\vspace{-0.1in}
\caption{FIXAR's operation sequence in a single timestep}
\label{scenario}
\vspace{-0.15in}
\end{figure}

Figure \ref{FIXAR} shows the overall architecture of FIXAR platform which broadly consists of the host CPU that emulates the RL environment and the FPGA accelerator that accelerates compute-intensive DNN operations of the actor and critic network. FIXAR is initiated by receiving its actor's current state and a random batch of $B$ transitions, i.e., a set of required input vectors including a state, an action, and a reward, from the environment run on CPU. The critic network evaluates Q-value of each transition and executes backward propagation (BP) and weight update (WU) based on the estimated Q-value. With updated weights, the critic network leads the BP and WU of the actor network in the direction where optimal actions can be obtained. Then, the actor network selects the action based on the updated weights in a given state and this forward propagation (FP) result is sent to host CPU. The environment takes the action computed from FPGA, calculates the reward, and changes to a new state. It stores transition information from the current step and samples a training batch in order to send to FPGA. In this way, FPGA continuously communicates with the host CPU through PCIe interface. Detailed operation sequence between the host CPU and FPGA accelerator is illustrated in Figure \ref{scenario}.
\par
FIXAR's FPGA accelerator is in charge of running all computationally heavy DNN inference and training workloads. It includes an activation memory that stores the input transitions from the host as well as the intermediate activations, a weight memory that stores model parameters of the actor and critic network, and a gradient memory that stores the intermediate gradient values. As the actor's model size (input:state-hidden:400-hidden:300-output:action) and the critic's model size (input:state+action-hidden:400-hidden:300-output:1) are relatively small, we are able to store all the model parameters in the weight memory, whose size is 1.05MB using only on-chip BRAMs. The size of the gradient memory is same as the weight memory's and it only uses on-chip BRAMs too. With accumulated gradient, weight update occurs in Adam optimizer module, which is fully local to FPGA as the entire model parameters are stored on-chip BRAMs. The size of activation memory is set to 2.94KB to hold the activation data out of all 3 layers. Since the accelerator has all the parameters and activations on-chip, it does not require any external DRAM memory accesses, which enables fast and efficient processing. For the compute side, the accelerator has multiple adaptive array processing cores where each of them contains 16x16 processing elements and an activation line buffer for data broad-casting. Weight data are pre-loaded from the weight memory. The outputs of the array cores are aggregated in the accumulator and passed to the activation unit for nonlinear functions. The pseudo random number generator (PRNG) module injects random noise to the final results of the actor's inference to help action exploration.

\section{\bf \textsc{Dynamic Fixed-Point Quantization of Reinforcement Learning}}
\vspace{-0.05in}

\begin{algorithm}[b]
\caption{Quantization-Aware Training for DRL}
\label{algorithm}
\begin{algorithmic}
\footnotesize
\setstretch{0.93}
\item
{\bf{Input:}} Quantization bit $n$,  Quantization delay $d$
\\
{\bf{Output:}} Trained reinforcement learning model $M$
\\
\\
Randomly initialize network parameter $\theta$ of $M$
\For{$t=1,T$}
\If{$t<d$}
\State{\bf{Activation: Fixed-point 32-bit, Weight: Fixed-point 32-bit}}
\State {Monitor the maximum and minimum value of}
\State{activations $A_{min}, A_{max}$}
\State{Update $\theta$ with activation $A$}
\Else 
\State{\bf{Activation: Fixed-point 16-bit, Weight: Fixed-point 32-bit}}
\State{$Q_n(A, A_{min}, A_{max}) = \lfloor{\frac{A}{\delta}}\rfloor+z$}
\State{where $\delta = \frac{|A_{min}| + |A_{max}|}{2^n} $ and $z=\lfloor{\frac{-A_{min}}{\delta}}\rfloor$}
\State{Update $\theta$ with $Q_n(A, A_{min}, A_{max})$}
\EndIf
\State{Evaluate(M)}
\EndFor
\end{algorithmic}
\end{algorithm}
FIXAR uses a dynamic fixed-point quantization for the first time, which changes the bit-precision of activations in the training process of deep reinforcement learning. To ensure the algorithmic accuracy measured by the level of accumulated reward, we adapt the Quantization-Aware Training (QAT) algorithm from QuaRL \cite{krishnan2019quantized} for a fixed-point version. In the original QAT algorithm, it learns the RL policy model based on the single-precision floating-point format for a certain time period defined as the quantization delay, then quantizes it to a narrow-bit floating-point representation for re-training. In our version, we start from the 32-bit fixed-point format and quantize it to half after a quantization delay. Algorithm \ref{algorithm} describes the fixed-point version of QAT algorithm used in the FIXAR platform.
\par
In the algorithm, the model parameter $\theta$ of both actor and critic network is initialized with random numbers. If the timestep $t$ is less than the quantization delay $d$, it performs BP and WU based on the input activations with 32-bit fixed-point format. During this time, the minimum and maximum value of the activations are actively monitored and captured. Once the timestep reaches the quantization delay, the previously captured minimum and maximum value are used to quantize the activations. From this point, activations are down-scaled to 16-bit fixed-point data and both BP and WU are done with quantized activations. Weights and gradients are kept in 32-bit fixed-point format for the entire timesteps. Because the weights are trained with full-precision activations for the quantization delay, a possible loss of accuracy can be compensated even we train them with half-precision activations for the remaining time. 
\renewcommand{\arraystretch}{1.2}
\vspace{-0.05in}
\section{\bf \textsc{Accelerator Design}}
\vspace{-0.01in}
The goal of the FIXAR's FPGA accelerator is twofold: to support DNN inference and training operations of the actor and critic network and to support dynamic fixed-point datapath for the proposed QAT algorithm. In this section, we mainly describe the design of adaptive array processing core, which is the main compute engine for the accelerator.
\vspace{-0.1in}
\subsection{Adaptive Array Processing Core}
\vspace{-0.03in}
As illustrated in Figure \ref{FIXAR}, the FPGA accelerator has $N$ adaptive array processing (AAP) cores for parallel DNN inference and training computations. Each core is arranged in a 2-dimensional array of configurable processing elements (PEs) where each PE performs multiply-and-accumulate (MAC) operation between activations and weights. In AAP core, a vector of input activations is copied from the activation memory to the 512-bit activation line buffer and the weights are pre-loaded to each PE from the weight memory. For operation, each activation is always broad-casted to a row of PEs while the partial sums from the PEs in a column are accumulated vertically to the accumulator at the bottom. After accumulated outputs are passed to the activation unit, they are saved in the activation memory and used as the input activations of the next layer.
\vspace{-0.03in}
\subsection{Dataflow for Adaptive Parallelism}
\vspace{-0.03in}
\begin{figure}[b]
\vspace{-0.2in}
\centering
   \includegraphics[width=7.5cm, height=5.2cm]{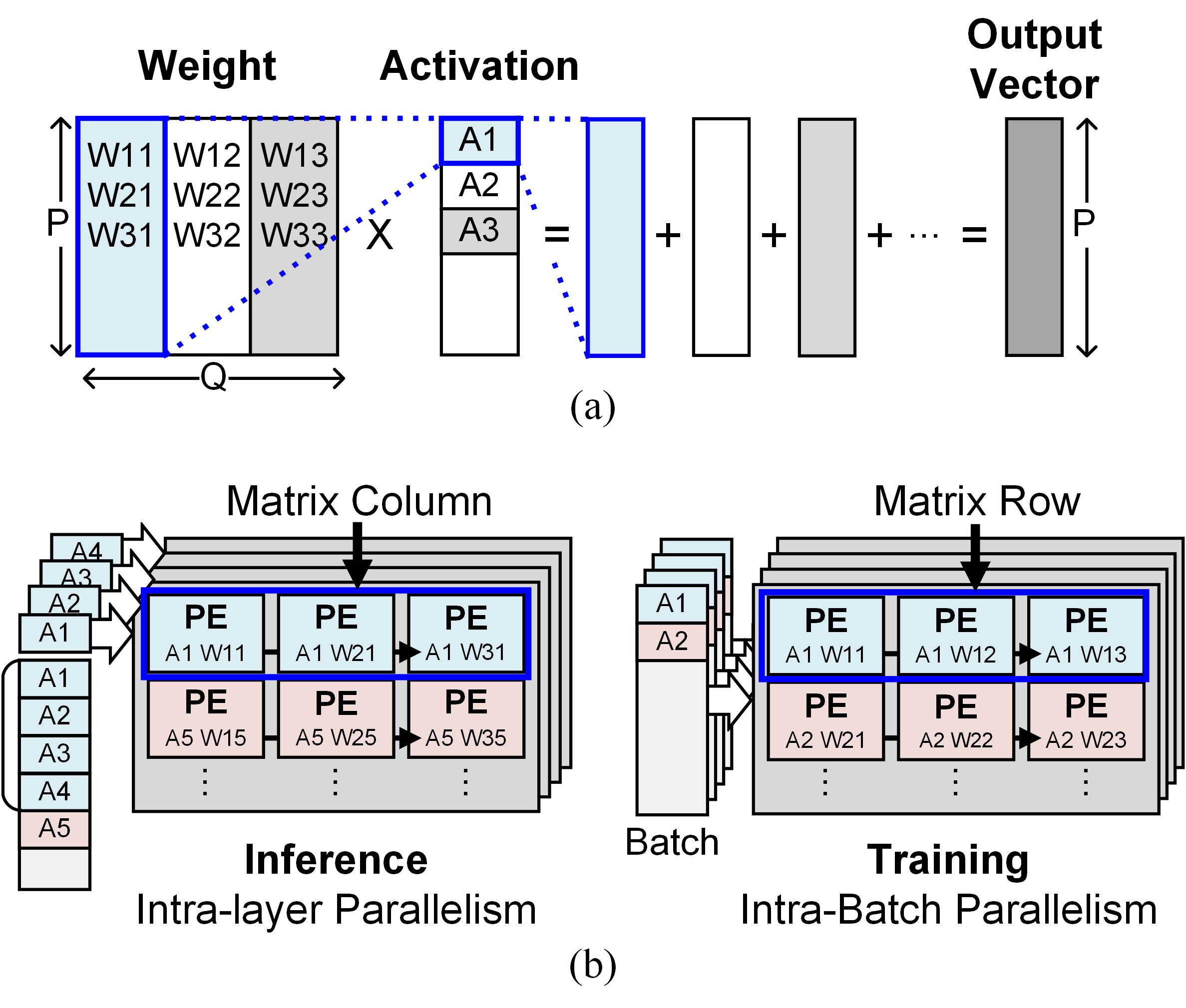}
   \hfil
\vspace{-0.1in}
\caption{(a) Column-wise matrix decomposition  (b) Dataflow for inference and training}
\label{columnwise}
\vspace{-0.1in}
\end{figure}
For each layer of a deep neural network, we need to calculate matrix-vector multiplication (MVM) for the weight matrix $W$ (size: PxQ) and an activation vector $A$ (size: Qx1). Among a few ways for MVM, we chose the column-wise matrix decomposition, as illustrated in Figure \ref{columnwise}(a). In this method, each column of the matrix is multiplied by the corresponding element of a vector (e.g., 1st column - 1st element, 2nd column - 2nd element, ..) to generate Q different partial-sum vectors. The output vector (size: Px1) can be calculated by accumulating all the partial-sum vectors. We use this mechanism for both forward-propagating inference and back-propagating training.
Figure \ref{columnwise}(b) shows the data mapping and flow for inference and training operation in AAP cores. For inference, we map each column of the matrix to a row of PE array and broad-cast an element of the vector to the row. Then, if we accumulate the products out of the PEs vertically, we can get the final vector at the bottom of the array. To have parallel processing within a single layer processing, i.e., intra-layer parallelism, we interleave each column of the matrix, which is mapped to each row of the PE array, among multiple AAP cores. For example, in the case of 4 AAP cores, the first core accumulates the partial-sum vectors from the 1st column, 5th column, and so on. In this case, once all the AAP cores finish their local accumulations, the results from the cores should be accumulated again to get the final vector. For training operation, we need a MVM with a transposed matrix $W^T$ and the error vector computed from the previous layer. Since we use the same column-wise mechanism in the training operation as well, we map each column of the transposed matrix to a row of PE array. Therefore, each row of the original matrix maps to a row of PE array. In training, we implement intra-batch parallelism by distributing the vectors in a batch across multiple AAP cores. As each AAP core works on each MVM, in parallel, we can increase the overall throughput by the number of AAP cores. Based on the adaptive parallelism enabled by different data mapping and flow, FIXAR is able to execute a single vector $N$ times faster in forward propagation and compute $N$ times more vectors in a batch in back-propagation without any off-chip DRAM access, where $N$ is the number of AAP cores.
\par
The weight memory stores the model's matrix parameters row by row over 16 BRAM modules. We set the bit-width of the weight memory to 512-bit to read or write 16 weights at the same cycle. For inference, the controller reads a single row from the weight memory and distributes them to a column of PE array. For training, on the other hand, it distributes the 16 weights to a row of PE array. Based on the column-wise matrix decomposition mechanism, we efficiently solve the matrix transpose problem, which happens in supporting both inference and training, by distributing the weights in a row of the matrix to a column or a row of the physical PE array. The weight memory is the centralized storage for model parameters and is shared among multiple AAP cores. Without any duplication in the weight memory, we are able to store the entire model parameters for the DDPG algorithm on-chip.
\begin{figure}[b]
\vspace{-0.2in}
\centering
   \includegraphics[width=8cm, height=3.3cm]{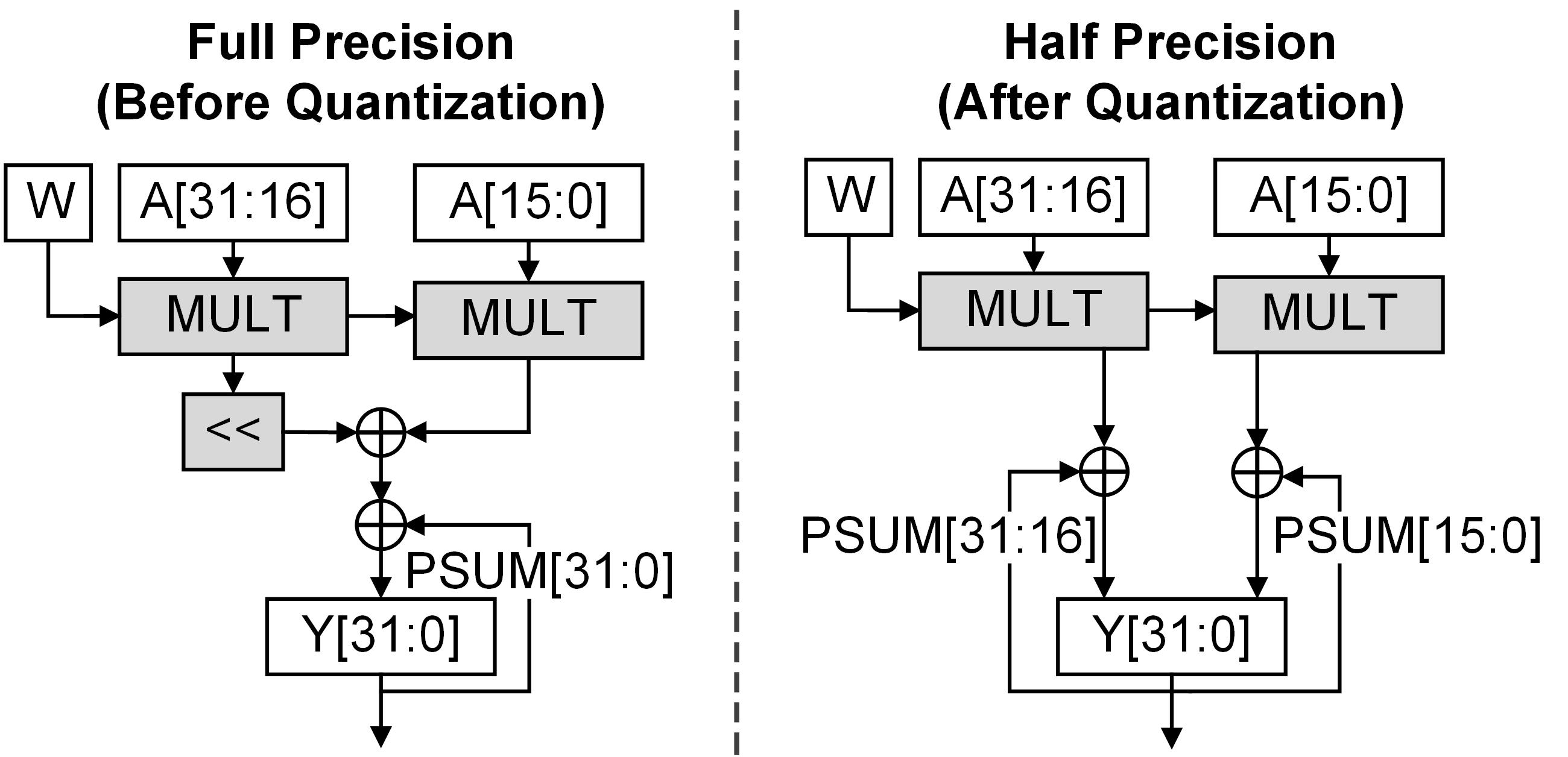}
   \hfil
\caption{Processing Element with Configurable Datapath}
\label{PE}
\end{figure}
\vspace{-0.08in}
\subsection{Processing element with configurable datapath}
\vspace{-0.03in}
Processing element (PE) of the AAP core is the most basic compute unit in the accelerator and performs multiply-and-accumulate (MAC) operation in fixed-point format. As each AAP core contains 256 PEs, having fixed-point arithmetic units instead of floating-point arithmetic units greatly reduces its logic area and power consumption. In accordance with the QAT, it supports the configurable datapath for 32-bit activation and two 16-bit activations, as shown in Figure \ref{PE}. Leveraging the fact that the 32-bit by 32-bit multiplication can be decomposed into two 32-bit by 16-bit multiplications, we employ two 32-bit by 16-bit multipliers in the PE. For the full-precision case before the quantization, the output from the upper-bits multiplier is left-shifted and added to the output of the lower-bits multiplier to generate the final result. For the half-precision case after the quantization, we use the two partial-sums separately as the final result. On the memory side, we don't expect any changes as two 16-bit activations just replace a single 32-bit activation. As a result, PEs with configurable datapath efficiently support both full and half-precision MACs in fixed-point and double the throughput for the half-precision case without any overhead.

\renewcommand{\arraystretch}{1.2}
\section{\bf \textsc{Evaluation}}
\vspace{-0.03in}
In this section, we evaluate the FIXAR's CPU-FPGA platform against the popular CPU-GPU platform. We used Intel Xeon 6226R at 2.9 GHz as the host CPU for both platforms. We used Xilinx Alveo U50 accelerator card for FPGA and Nvidia Titan RTX for GPU.
\vspace{-0.05in}
\subsection{Implementation Results}
\vspace{-0.05in}

\begin{figure}[t!]
\centering
   \includegraphics[width=4cm]{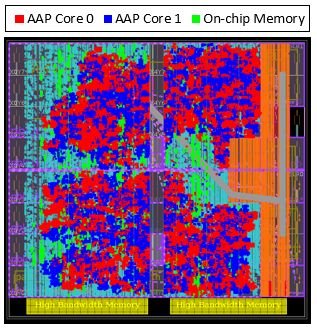}
   \hfil
\caption{FPGA layout}
\label{board}
\vspace{-0.05in}
\end{figure}

\begin{table}[t]
\centering
\caption{FPGA Resource Usage on Xilinx Alveo U50}
\vspace{-0.05in}
\label{Utilization}
{\footnotesize
    \begin{tabular}{m{7em} p{3em} p{3em} p{3em} p{3em} p{3em}}
        \toprule
        {\bf Component} & {\bf LUT} & {\bf FF} & {\bf BRAM} &{\bf URAM} & {\bf DSP}\\
        \hline\hline
        PEs &  216.3K & 161.8K & 0 & 0 & 2295\\
        \hline
        On-chip Memory & 10.3K & 0 & 584 & 128 & 0 \\
        \hline
        Adam Optimizer & 46.7K & 70.2K & 0 & 0 & 3 \\
        \hline
        Control Unit & 69.0K & 45.4K & 0 & 0 & 0\\
        \hline
        Kernel Interface & 68.8K & 15.2K & 12 & 0 & 0\\
        \hline
        HBM Interface & 8.2K & 13.1K & 2 & 0 & 0 \\
        \hline
        PCIe DMA & 88.8K & 103.2K & 176 & 0 & 4 \\
        \hline
        \hline
        Total & 508.1K & 408.8K & 774 & 128 & 2302\\
         & (58.4\%) & (23.5\%) & (57.6\%) & (20.0\%) & (38.8\%)\\
        \bottomrule
    \end{tabular}
}
\vspace{-0.2in}
\end{table}
We implement the FIXAR's hardware accelerator on the Xilinx Alveo U50 acceleration card. As the U50's FPGA chip uses chiplet-based design with 2 super logic regions (SLRs), we carefully designed the pipelining in the PE array and the fan-out of the weight and gradient memories to overcome a severe SLR crossing penalty. As a result, we are able to integrate 2 AAP cores across the 2 SLRs at 164MHz operating frequency with utilizing 58.4\%, 57.6\%, and 38.8\% of LUT, BRAM, and DSP resource, respectively. We used Xilinx Vitis framework 2020.1 for PCIe communication between the host and FPGA card. Figure \ref{board} shows the final layout of Xilinx U50 FPGA and Table \ref{Utilization} summarizes its resource usage.

\begin{figure}[t]
\vspace{-0.2in}
\centering
   \includegraphics[width=7cm, height=3.9cm]{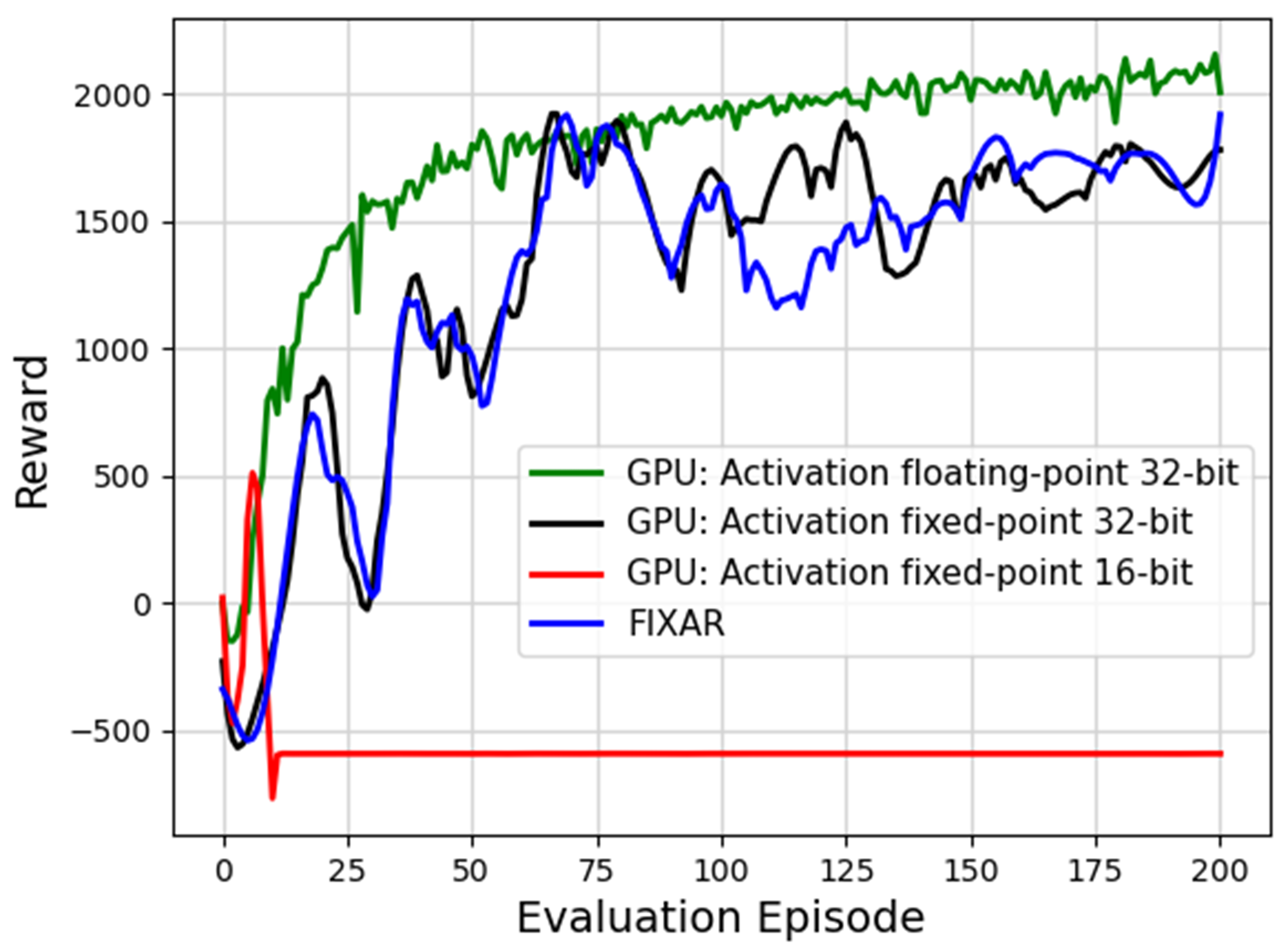}
   \hfil
\caption{Algorithm accuracy in \textit{HalfCheetah} environment}
\label{accuracy}
\vspace{-0.15in}
\end{figure}

\vspace{-0.05in}
\subsection{Benchmarks for Continuous Action Space}
To evaluate the FIXAR platform, we run multiple physical locomotion benchmarks \textit{HalfCheetah}, \textit{Hopper}, and \textit{Swimmer} from MuJoCo physics engine \cite{6386109}. These benchmarks target continuous action spaces and are considered as complex tasks, hence they are widely used for RL algorithm evaluation. For instance, the \textit{HalfCheetah} benchmark aims to train a cheetah to run by giving 6 action outputs based on the cheetah's state including 17 physical conditions and the reward from the environment. Likewise, \textit{Hopper} benchmark has 11-dimensional state and 6-dimensional action, and \textit{Swimmer} benchmark has 8-dimensional state and 2-dimensional action. We use DDPG algorithm to learn agent's action policy in the continuous action space. In our DDPG implementation, we use a neural network with 2 hidden layers (input:state-hidden:400-hidden:300-output:action) for the actor. The critic's neural network receives both the state and action as inputs and generates a single error value to the actor (input:state+action-hidden:400-hidden:300-output:1). Both networks use rectified linear unit (ReLU) in each layer while the actor applies additional tanh to the output. Both network parameters are optimized using Adam optimizer with a learning rate of $10^{-4}$. We run the task for 1 million timesteps in total while evaluating the reward every 5000 timesteps. For each evaluation, we calculate the average of cumulative rewards until the agent falls down for given random 10 states.
\vspace{-0.05in}
\subsection{Performance}
\vspace{-0.03in}
{\bf Algorithm accuracy} To verify the algorithm accuracy of the FIXAR's fixed-point training, we measure the total reward of it for training episode (episode = 1000 timesteps). For comparison, we also measure the training results of GPU when it runs the experiment for various data formats including 32-bit single-precision floating-point, 32-bit fixed-point, and 16-bit fixed-point. The graph in Figure \ref{accuracy} presents the total reward obtained during the training process. It shows that the FIXAR's dynamic fixed-point successfully trains the DDPG DNNs with the cumulative reward level saturating towards 2000, like in the 32-bit floating-point and 32-bit fixed-point. Although FIXAR has a dip in reward after quantization, it scoops up again as it re-trains the model with a reduced bit-precision. This is possible because it starts re-training from the pre-trained model with a full bit-precision during the quantization delay time before the quantization happens. On the other hand, the GPU case that starts the training with the 16-bit fixed-point fails to train.
\par
\begin{figure}[t]
\vspace{-0.2in}
\centering
   \includegraphics[width=8.5cm, height=4cm]{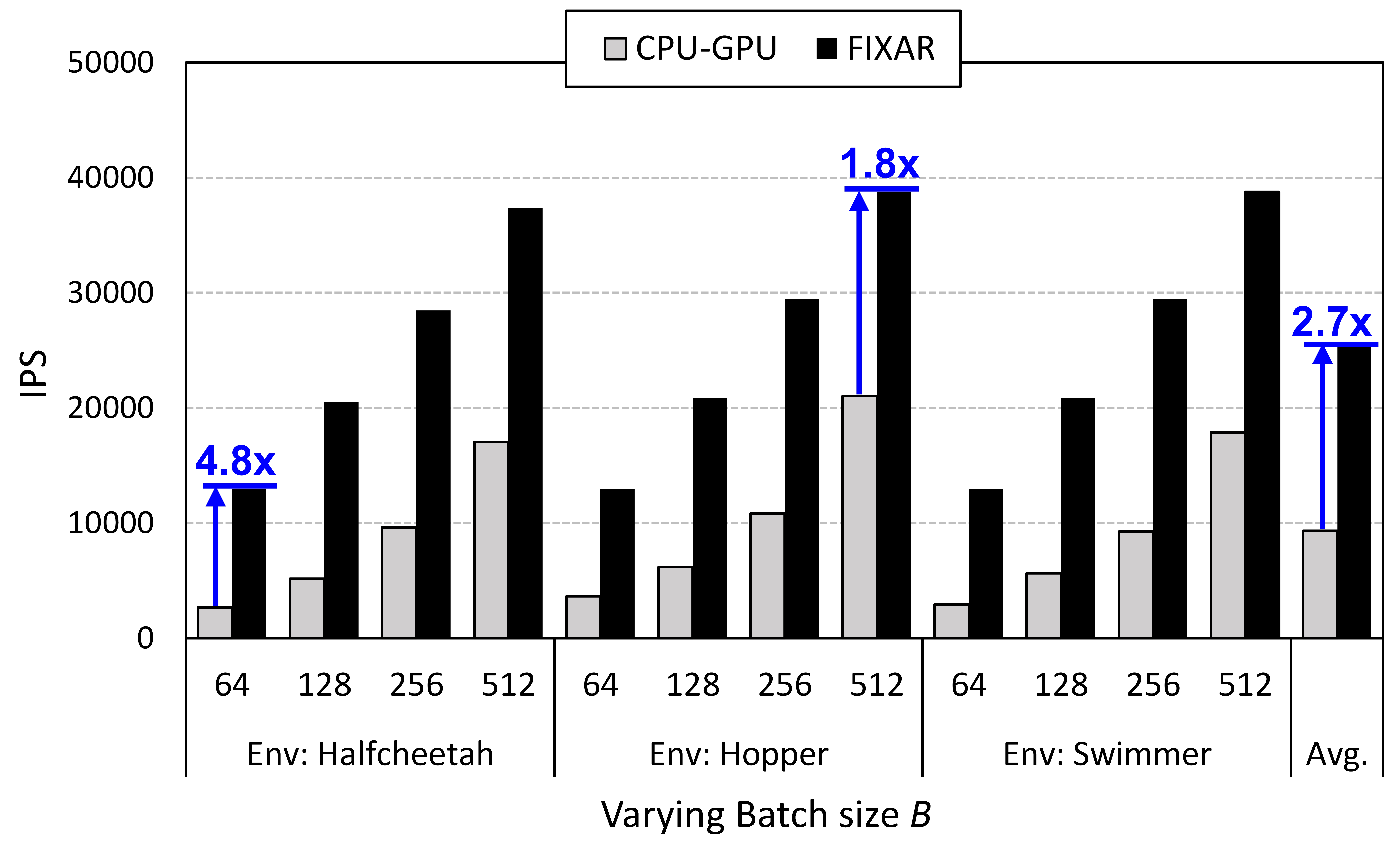}
   \hfil
\vspace{-0.07in}
\caption{FIXAR platform's training throughput}
\label{result}
\end{figure}
\begin{figure}[t]
\centering
   \includegraphics[width=8.5cm, height=4.1cm]{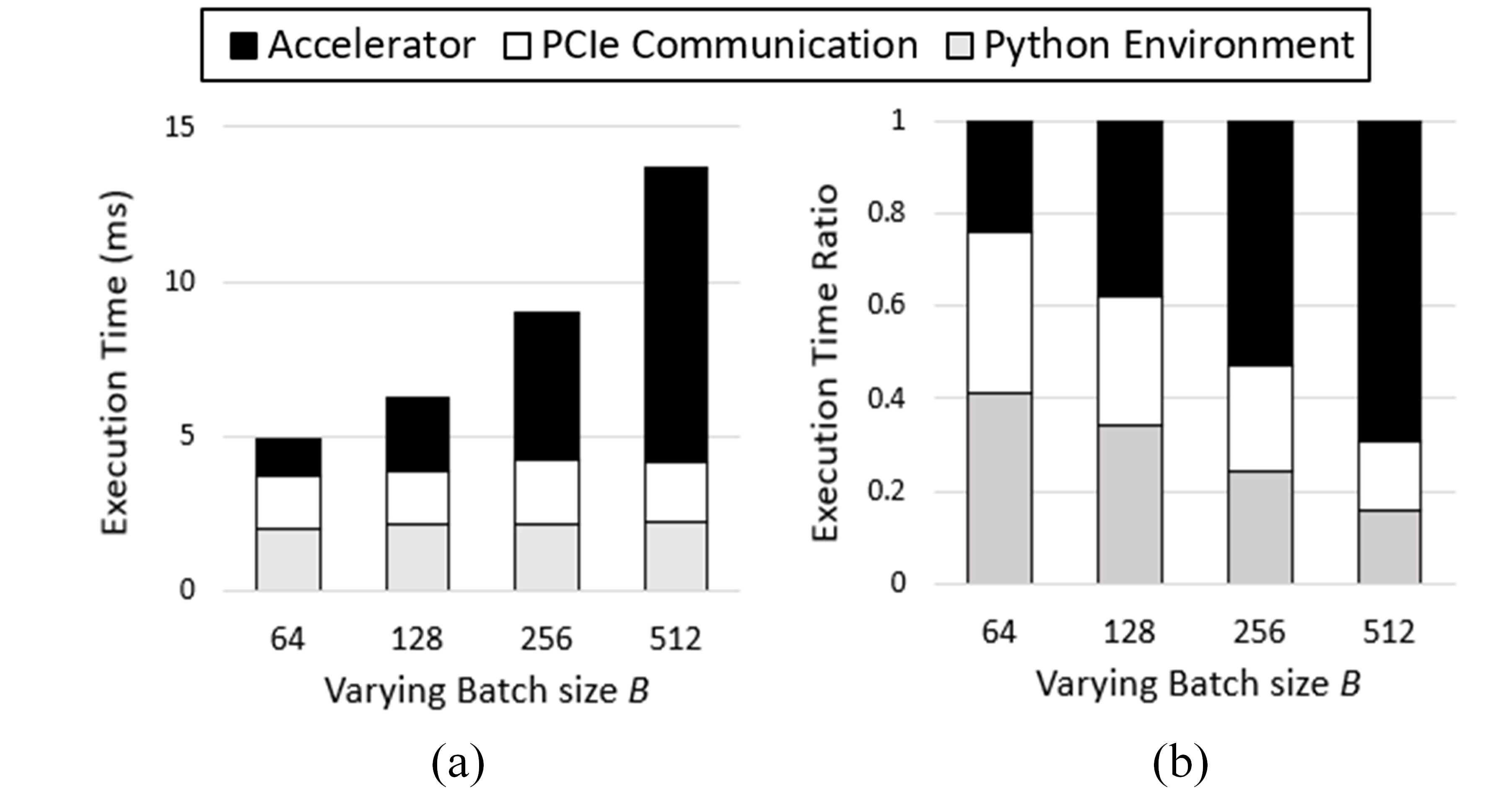}
   \hfil
\vspace{-0.07in}
\caption{(a) Execution time of FIXAR platform (b) Execution time ratio of FIXAR platform}
\label{runtime}
\vspace{-0.2in}
\end{figure}

{\bf Training throughput} In accordance with previous works, we use the metric called IPS, the number of inferences processed per second, to evaluate the training performance. It is the ratio of the total number of collected samples to the end-to-end system time taken in an entire timestep including inference, training, and environment interactions.
Figure \ref{result} shows the IPS results of the FIXAR and CPU-GPU platform for the chosen benchmarks when the batch size varies among 64, 128, 256, and 512. As the batch size increases, the throughputs of both platforms improve as well. Figure \ref{runtime}(a) shows the detailed execution time breakdown of a single timestep in the FIXAR platform for different batch sizes. For all cases, the CPU time spent on running MuJoCo Python environment is rather constant around 2 ms. The time spent on Xilinx run-time to import the input batch from the host CPU to the FPGA increases marginally even though the batch size doubles up. This tells us that the initial overhead of buffer allocation and PCIe communication in the run-time is quite large. The time spent on FPGA accelerator is linear to the size of input batch because the FIXAR's AAP cores remain highly utilized for different batch sizes thanks to the intra-batch parallelism. We can also observe that the system bottleneck changes from the CPU to the FPGA accelerator as the batch size increases, as shown in Figure \ref{runtime}(b). The CPU-GPU platform benefits from a large batch size more than FIXAR because it helps GPU improve the utilization significantly. As a result, the FIXAR platform performs 1.8-4.8 times better than the CPU-GPU platform although it has some inefficiency in the run-time system.
\par

\begin{figure}[b!]
\vspace{-0.1in}
\centering
   \includegraphics[width=8.5cm, height=7.5cm]{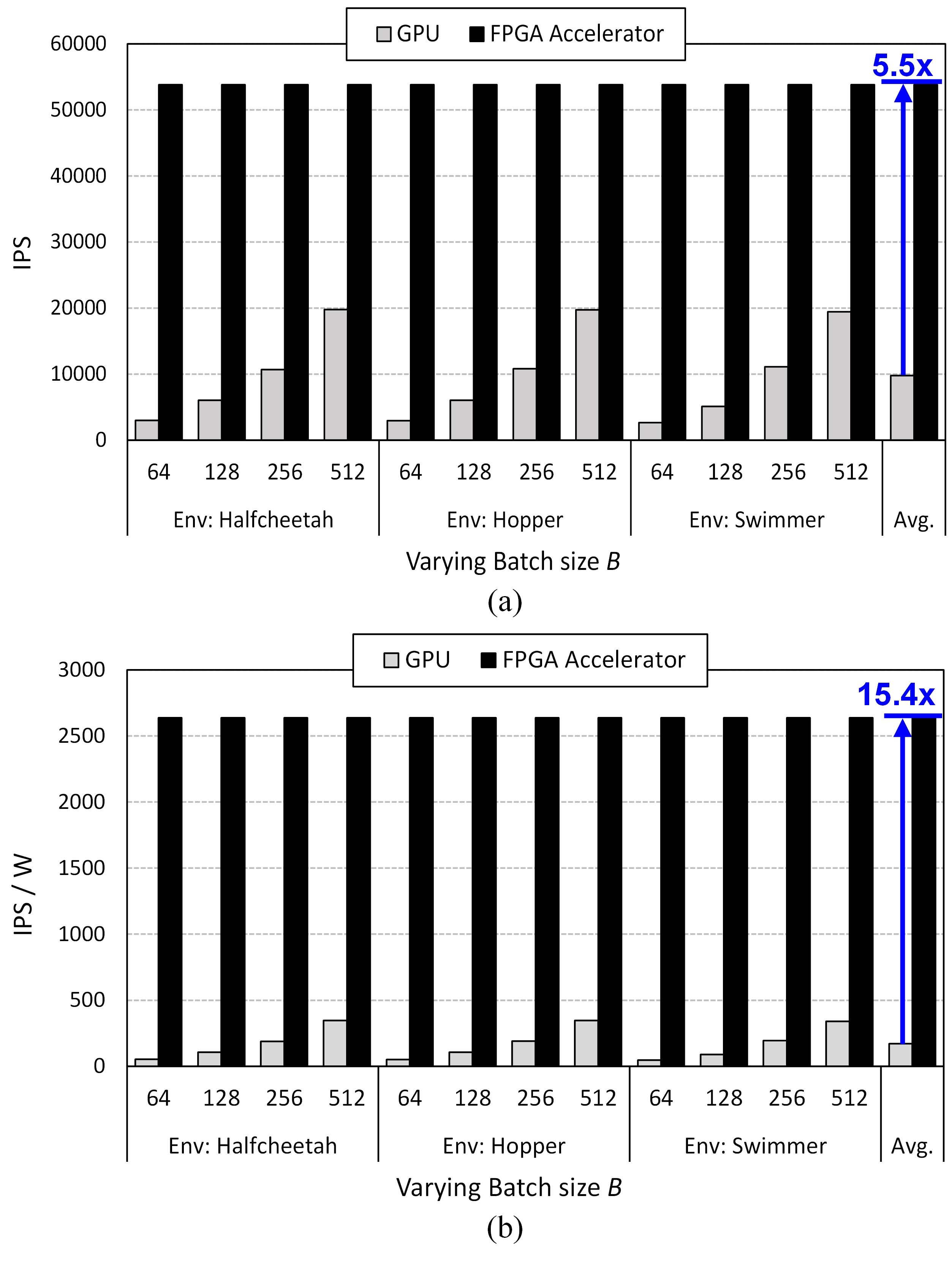}
   \hfil
\vspace{-0.1in}
\caption{(a) Accelerator throughput (b) Accelerator energy efficiency}
\label{accelerator}
\end{figure}
{\bf Accelerator efficiency}
Figure \ref{accelerator} shows the IPS throughput performance and their energy efficiency measured only on the accelerators, i.e., FPGA accelerator and GPU, except the host CPU time. The IPS performance of the FIXAR's accelerator remains high at 53826.8 IPS for all different batch sizes. This is possible because the multiple AAP cores (N=2 in this implementation) can run a single vector faster using the intra-layer parallelism in forward path as well as can run multiple vectors in a batch in parallel using the intra-batch parallelism in backward path as described in Section 5-B. Since the batch size is big enough compared to the number of AAP cores, the hardware utilization is kept very high at 92.4\%. On the other hand, GPU's hardware utilization linearly increases as the batch size increases. For power estimation, we used Xilinx Board Utility that measures the power consumption of the overall acceleration card including FPGA, PCIe interface, and on-board DRAMs. We measured that the FPGA and GPU consume 20.4W and 56.7W power on average, respectively, for running DNN models of the DDPG algorithm with 3 benchmarks. As a result, FIXAR's FPGA accelerator showing 2638.0 IPS/W energy efficiency achieves 15.4 times higher energy efficiency than that of Titan RTX GPU.
\vspace{0.05in}

\section{\bf \textsc{Related Work}}
\vspace{-0.07in}
\begin{table}[t!]
\centering
\caption{Comparison Table with Previous Works}
\label{comparison table}
{\scriptsize
    \begin{tabular}{m{7.5em} p{7.5em} p{6.5em} p{6.5em}}

    \toprule
         & {\bf ASPLOS'19 [19]} & {\bf FCCM'20 [20]} & {\bf FIXAR}\\
        \hline\hline
        Platform & Xilinx VCU1525 & Xilinx U200 & Xilinx U50 \\
        \hline
        Clock &  180MHz & 285MHz & 164MHz  \\
        \hline
        Algorithm & Actor-Critic (A3C) & Actor-Critic (PPO) & Actor-Critic (DDPG)\\
        \hline
        Task Env. & Discrete & Continuous & Continuous \\
        \hline
        Precision & Floating 32-bit & Floating 32-bit & Fixed 32, 16-bit\\
        \hline
        DSP & 2348 & 3744 & 2302\\
        \hline
        Network Size & 2592.0 KB & 229.6 KB & 514.4 KB\\
        \hline
        Peak Perf. & 2550.0 IPS & 15286.8 IPS & 38779.8 IPS\\
        \hline
        Normalized Peak Perf. to FIXAR & 12849.1 IPS & 6823.2 IPS & 38779.8 IPS\\
        \hline
        Energy Efficiency (Accelerator) & 141.7 IPS/W & - & 2638.0 IPS/W\\
        \bottomrule
    \end{tabular}
}
\vspace{-0.15in}
\end{table}
There have been increasing number of works on accelerating deep reinforcement learning over the last few years, but most of them implemented table-based Q-learning algorithms\cite{article}. A couple of recent works focused on accelerating advanced actor-critic algorithms like FIXAR. FA3C \cite{cho2019fa3c} presented a framework for accelerating a deep reinforcement learning algorithm named A3C\cite{mnih2016asynchronous} and applied it for an Atari game. However, FA3C was only evaluated in the discrete action space which requires a much lower precision than the continuous action space that FIXAR targets. In \cite{9114846}, a FPGA based acceleration platform is proposed for PPO algorithm \cite{DBLP:journals/corr/SchulmanWDRK17}. While the PPO accelerator targets the continuous action space like FIXAR, it is still based on resource-hungry floating-point format. In addition, its DNN model size is less than half of the FIXAR's. Based on efficient fixed-point arithmetic units, FIXAR is able to process more operations with less DSPs so its overall throughput surpasses even with a more-than-twice larger network. Table \ref{comparison table} summarizes the comparison of the FIXAR platform against the state-of-the-art works.
\par

\vspace{0.03in}
\section{\bf \textsc{Conclusion}}
\vspace{-0.03in}
In this paper, we present a DRL acceleration platform called
FIXAR, which employs fixed-point data types and arithmetic units for the first time. We propose the quantization-aware training that enables to reduce the data precision in fixed-point by half after a certain number of training steps. We also design a FPGA accelerator that employs adaptive dataflow and parallelism to handle both inference and training operations with supporting dual fixed-point data types. We evaluate the implemented FIXAR platform against the conventional CPU-GPU platform by running multiple benchmarks for continuous action spaces. As a result, FIXAR achieves 25293.3 IPS training throughput and 2638.0 IPS/W accelerator efficiency, which are 2.7 times higher and 15.4 times more energy efficient than those of the CPU-GPU platform. It also shows the best performance and energy efficiency among other acceleration platforms even it runs one of the most complex DNN models thanks to its fixed-point arithmetic implementation.
\vspace{-0.05in}

\begingroup
\setstretch{0.91}
\bibliographystyle{IEEEtran}
\bibliography{IEEEabrv, 8_reference}
\endgroup

\end{document}